# Recent results in neutrino physics

D.V. Naumov
Joint Institute for Nuclear Research
Dubna, Russia
dnaumov@jinr.ru

ABSTRACT

This manuscript is a shorthand version of my talk given at Odessa Gamov School on Astronomy, Cosmology and Beyond (22-28 August 2011, Odessa, Ukraine). Within this note we very briefly review the main achievements, new results and open problems in neutrino physics of today.

## 1. Introduction

Neutrino is a light spin one-half fermion participating in weak and gravitational interactions. Its history begins with a letter of W.Pauli (4$^{th}$ December 1930) to Lise Meitner et al. in which he suggested an existence in nuclei of electrically neutral particles with a small mass which he called «neutron». With help of these «neutrons» Pauli attempted to solve several puzzles: seemingly violation of energy-momentum conservation in $\beta$-decays (continuus $\beta$-spectrum instead of expected discrete spectrum) and «wrong» spin statistics of nuclei $^6$Li, $^{14}$N (these nuclei have an integer magnetic spin which could not be composed of an odd number of spin one-half protons). The first theory of weak interactions was given by E.Fermi who also introduced a new name for a light neutral particle participating in these interactions — «neutrino». It is probably worth to mention that modern solution to Pauli's problems of 1930 actually requires two different particles: *neutron and neutrino* while this fact is usually omitted refering to Pauli's letter as to a theoretical discovery of neutrino.

### 1.1 Number of types of neutrino

Experimental discovery of three neutrino flavours span in time for more than 50 years: 1956 ($\nu_e$), 1962($\nu_\mu$) and 2000 ($\nu_\tau$). The flavour of neutrino is not however a good quantum number — it was experimentally found that it is not conserved. Instead it is better to think about number of neutrino species — number of particles with different masses. Assuming universality of weak interactions, Z boson decays by LEP experiments constraint this number as $N_\nu = 2.9841 \pm 0.0083$. However these data do not exclude existence of either heavy neutrinos (with mass heavier than $m_Z/2$) or of neutrino with non standard interactions, like *sterile neutrino*. Remarkably the number of neutrino types following from analysis of cosmological data $N_\nu = 4.3 \pm 0.9$ while being in agreement with LEP number within the error bars has some tention with accelerator data.

### 1.2 Neutrino mass and mixing matrix

What do we know about mass of neutrino? Direct measurements based on analysis of kinematics give the following constraints [1]:

$$\sum_i |V_{ei}|^2 m_i < 2.3\,\text{eV}, \quad \sum_i |V_{\mu i}|^2 m_i < 170\,\text{keV}, \quad \sum_i |V_{\tau i}|^2 m_i < 15.5\,\text{MeV},$$

where the most stringent limit comes from analysis of tritium decays. Cosmology puts another limit of sum of masses of neutrinos [2]:

$$\sum_i m_i < 1-2\,\text{eV}.$$

These data do not exclude zero mass of neutrino. There is however an important piece of information which tells us that at least two of three neutrino are massive. This conclusion comes from the analysis of neutrino oscillations — a macroscopic display of quantum mechanical interference.

These data give the following: $m_2^2 - m_1^2 = (7.65 \pm 0.23) \cdot 10^{-5}\, eV^2$, $|m_3^2 - m_1^2| = (2.4 \pm 0.12) \cdot 10^{-3}\, eV^2$.

From these data and combining it with cosmological limit it is easy to conclude that mass of the most heavy neutrino should be at least larger than $\sqrt{|m_3^2 - m_1^2|}$ and lighter than sum of masses of all neutrinos. Thus the mass of the heaviest neutrino is bound within a rather narrow window $0.05 < m_\nu^{heavy} < 1-2\,\text{eV}$.

It appears that every massive neutrino $\nu_i$ interacts with every charged lepton $l_\alpha^\pm$ proportionally to the corresponding element $V_{\alpha i}$ of the Pontecorvo-Maki-Nakagawa–Sakata (PMNS) neutrino mixing matrix:

$$V_{\alpha i}=\begin{pmatrix} V_{e1} V_{e2} V_{e3} \\ V_{\mu 1} V_{\mu 2} V_{\mu 3} \\ V_{\tau 1} V_{\tau 2} V_{\tau 3} \end{pmatrix}=\begin{pmatrix} c_{12}c_{13} & s_{12}c_{13} & s_{13}e^{-\delta} \\ -s_{12}c_{23}-c_{12}s_{23}s_{13}e^{\delta} & c_{12}c_{23}-s_{12}s_{23}s_{13}e^{i\delta} & s_{23}c_{13} \\ s_{12}c_{23}-c_{12}s_{23}s_{13}e^{\delta} & -c_{12}c_{23}-s_{12}s_{23}s_{13}e^{i\delta} & c_{23}c_{13} \end{pmatrix} \quad (1)$$

where $\alpha=e,\mu,\tau$ while $i$ runs from 1 to 3 and in (1) we used the following abbreviations $c_{ij}=\cos\theta_{ij}, s_{ij}=\sin\theta_{ij}$. Some elements of the neutrino PMNS matrix (1) has been measured in experiments with solar, atmospheric and reactor neutrinos.

The following matrix elements are known today: $\sin^2 2\theta_{12}=0.30^{+0.02}_{-0.02}, \sin^2 2\theta_{23}=0.50^{+0.07}_{-0.06}$.

### 1.3 Missing angles $\theta_{13}$ and $\delta$ in the mixing matrix

There are two angles of PMNS matrix still unknown: $\theta_{13}$ and $\delta$. Measurement of these missing angles is one of the main directions of current research in neutrino physics. $\delta$ - is the phase parameter responsible for CP-violation in lepton sector. It is crucial to measure this parameter as it may shed light on baryo- and leptogenesis in the early Universe. It is however impossible to measure $\delta$ if $\theta_{13}=0$ and this largely explains an interest of the community to measure $\theta_{13}$. What do we know about $\theta_{13}$ today? A number of experiments constrained it from above.

$\sin^2 2\theta_{13}<0.17$ limit was obtained by Chooz experiment [3] and similar limit was given by Palo Verde [4] (both reactor experiments). $\sin^2 2\theta_{13}<0.26$ limit was given by K2K [5] and the following limit $\sin^2 2\theta_{13}<0.15$ was given by MINOS [6] both being accelerator experiments.

An indication on non-zero value of $\theta_{13}$ came from reactor experiment KamLAND [7]: $\sin^2 2\theta_{13}=0.02^{+0.016}_{-0.016}$ which being combined with other world data yields the best fit value $\sin^2 2\theta_{13}=0.009^{+0.013}_{-0.007}$ [8].

In summer of 2011 T2K Collaboration and MINOS Collaboration claimed an experimental evidence for non-zero value of $\theta_{13}$. T2K in appearance mode claimed to observe an excess of 6 $\nu_e$ candidates with an expected background of 1.5 [9]. If interpreted as $\nu_\mu \to \nu_e$ oscillations this implies quite large $\sin^2 2\theta_{13}$ = 0.1±0.07. Its statistical significance is 2.5σ which is not enough to be a «discovery» yet. MINOS also in appearance mode claimed to observe a small excess of $\nu_e$ events with statistical significance of about 1.7σ which is well compatible with a fluctuation [10]. However in a global analysis of neutrino oscillation world data [11] $\theta_{13}$ is non-zero at 3σ confidence level. The best fit value is $\sin^2 2\theta_{13}$ = 0.02. Today a measurement of $\theta_{13}$ is a hot topic in neutrino physics and several experiments with anti-neutrinos from reactors (Double Chooz, Reno, DayaBay) and with neutrinos from accelerator (T2K, Nova, MINOS) are addressing this topic and competing with each other.

### 1.4 Magnetic moment of neutrino

Does neutrino possess a magnetic moment? The Standard Model (SM) predicts it to be unobservably small on the level of (for heaviest neutrino):
$$\mu_\nu=3eG_F m_\nu/(8\pi 2\sqrt{2})=3.2\cdot 10^{-19} m_\nu/eV \approx 10^{-20} \mu_B.$$
The experiments so far could put only upper limits. The current limit $\mu_\nu<5.4\cdot 10^{-11} \mu_B$ is given by Borexino Collaboration [12]. More stringent limits come from GEMMA and GEMMA-2 reactor experiment: $\mu_\nu<5\cdot 10^{-12}-3.2\cdot 10^{-11} \mu_B$ [13]. However these data is not used by PDG Collaboration.

### 1.5 Lifetime of neutrino

Is neutrino a stable particle? In the SM the answer is obvious - neutrino does not decay. However experimental limits are surprisingly modest compared to limit on the proton life-time ($\tau_p>10^{38}-10^{40}$ years). The most stringent limits on neutrino life-time come from measurements of its magnetic moment. An analysis of solar neutrino data yields: $\tau_\nu/m_\nu>7\cdot 10^9 s/eV$, which implies $\tau_\nu/m_\nu>2.8\cdot 10^8 s$ for heaviest ν.

### 1.6 Dirac or Majorana?

One of most important questions in neutrino physics — is neutrino a Dirac or Majorana particle? In other words if neutrino and anti-neutrino are two different particles (Dirac) or this is the same particle (Majorana)? Naively, one might think that since so far it was never observed experimentally that anti-neutrino can cause a reaction like:

$$\bar{\nu}_e + n \rightarrow p + e \quad (2)$$

which is caused by neutrino

$$\nu_e + n \rightarrow p + e \quad (3)$$

than this might indicate in favour that neutrino is a Dirac fermion. In fact, this is not a proof because of V-A type of weak interaction in the SM which favours left-handed helicty of neutrino and right-handed helicty of antineutrino. Therefore, even if neutrino is a Majorana particle than the probability of reaction (2) with $\nu_e = \bar{\nu}_e$ emitted from (for example) $n \rightarrow p + e + \bar{\nu}_e$ decay will be dramatically suppressed by the factor of the order of $m_\nu^2 / E_\nu^2 \approx 10^{-22}$ (at neutrino energy $E_\nu = 1$ GeV) due to opposite helicities of neutrino in the initial and final states of these reactions. This is hard to detect in conventional experiments with (anti)neutrino beam scattering off the target. One of the most promising technique to investigate the neutrino nature (Dirac or Majorana) is to observe neutrinoless double beta decay $0\nu2\beta$ of heavy nuclei $(A, Z) \rightarrow (A, Z-2) + 2e^-$. This reaction is only possible if neutrino is a Majorana particle. This method is sensitive to neutrino of light mass of the order of eV.

A number of experiments aimed to address this issue [14,15]. Some future project are under preparations [16]. The experiments use various nuclei to probe the neutrino nature. So far there is no solid evidence in favour of Majorana neutrino and the experiments could put only limits on life-time of exploited nucleus against $0\nu2\beta$ decay. Using thus obtained limits one could put a limit on neutrino mass combination $m_{eff}^2 = \left| \sum_i V_{ei}^2 m_i \right|^2$. As an example let us mention the limits following from NEMO-3 experiment: $T_{1/2}^{0\nu2\beta} > 1.8 \times 10^{22}$ years at 90% CL which implies $m_{eff}^2 < 4.0 - 6.3$ eV².

For heavy Majorana neutrino (for the masses of the order of TeV) $0\nu2\beta$ process becomes impractical. There is another type of reaction which could probe nature of neutrino with TeV range mass: $l^- l^- \rightarrow W^- W^-$ at colliders [17]. This reaction is also possible only if neutrino is Majorana particle.

## 2. Recent results

Last several years are highlighted by a number of new, interesting and sometimes very unexpected results obtained in neutrino physics. Let us briefly recall some of these results.

### 2.1 Solar neutrino puzzle

A long standing puzzle of solar neutrinos was finally solved after measurements of the SNO [18], SuperKamikande [19], KamLAND [20] and Borexino Collaborations [21]. SNO accurately measured the number of neutrino scatterings off heavy water (D$_2$0) due to both charged (CC) and neutral (NC) currents. These measurements are sensitive to $\nu_e$ (CC) and $\nu_e + \nu_\mu + \nu_\tau$ (NC) fluxes thus unambigiously pointing if neutrino oscillation occured or not. A global analysis of the data provided by the above mentioned experiments ensures us that the solar neutrino puzzle is solved due to neutrino oscillations.

### 2.2 Reactor anomaly

The KamLAND Collaboration experimented with reactor $\bar{\nu}_e$ searching for their disappearence. The japanese reactors contributing to the KamLAND data have on average a distance of about 180 km from the detector. The typical energy of $\bar{\nu}_e$ from reactors (several MeV) and distance between source and detector of about 200 km turned out to be an ideal combination of parameters in order to observe $\bar{\nu}_e$ oscillations. All previous attempts to observe $\bar{\nu}_e$ oscillations with reactors were not successful just because of too short distance between reactor and detector used in the previous experiments. Therefore, the reactor experiments with short base were in a good agreeement with theoretical expectations assuming no neutrino oscillations.

However in the beging of 2011 this agreement was seriously questioned after a new detailed theoretical calculation of $\bar{\nu}_e$ fluxes from reactors [22]. The new fluxes are predicted to be by 3% larger than previous estimates which makes now a tension with the world reactor data. Some physicists interpret this discrepancy as a manifistation of new neutrino state — *sterile neutrino*. This possibility is still an open question in neutrino physics.

## 2.3 Geoneutrino

Nowdays a well educated schoolchild knows that interior of the Earth is quite hot and is hotter towards the center of our planet. However why it is like that — nobody can tell for sure. There are various hypotheses and we could not discuss all of them in details here. The most popular however are the following three:
- The heat of the Earth interior still remains after the primary heating of the proto-planet.
- Weak decays of radioactive nuclei like $^{238}$U, $^{232}$Th, $^{40}$K during billions of years continuously heat the interior of the planet.
- There is a sort of an active nuclear reactor in the center of the Earth as a source of the internal energy (georeactor).

Perhaps all these mechanisms contribute to some extent. The last two mechanisms should produce $\bar{\nu}_e$ which could be detected on the Earth surface by neutrino experiments. Two neutrino experiments KamLAND [23] and Borexino [24] searched for such geoneutrinos and found an evidence for them with a combined statistical significance of 4.2σ. However the sensitivity reached by both experiments is not enough to prefer any among the above mentioned three models, while one could put a limit on the possible power of georeactor. Its power should not be larger than 3GW. The next progress in this interesting field is expected when new detectors will be functional: Hanna-Hanna, Lena and SNO+.

## 2.3 Atmospheric neutrino puzzle

Another puzzle in neutrino physics was related to the so called atmospheric neutrinos — particles produced in decays of hadrons and leptons which in their turn are produced in interactions of cosmic rays with nuclei in the atmosphere. Qualitatively one could expect the number of muon neutrinos and antineutrinos to be twice of that of electron neutrinos and antineutrinos because of the following chain of reactions:
$\pi^+ \to \mu^+ + \nu_\mu, \mu^+ \to e^+ \nu_e \bar{\nu}_\mu$ and $\pi^- \to \mu^- + \bar{\nu}_\mu, \mu^- \to e^- \bar{\nu}_e \nu_\mu$. The SuperKamikande however observed these numbers to be nearly the same which, among other possible interpretations, could be interpreted as a result of $\nu_\mu \to \nu_\tau$ oscillations.

The MINOS Collaboration significantly improved the previous measurement of SuperKamiokande Collaboration of $(|m_3^2 - m_1^2|, \sin^2\theta_{23})$ parameter space also favouring neutrino oscillations as a solution to atmospheric neutrino problem [25].

In 2010 MINOS reported also a hint for possible difference between $(|m_3^2 - m_1^2|, \sin^2\theta_{23})$ of neutrino and anti-neutrino which would mean CPT violation. However one year later a new analysis of MINOS did not confirm the previous hint. The MinoBooNE Collaboration reported to observe neutrino oscillations in $\bar{\nu}_e \to \bar{\nu}_\mu$ channel while there is no hint for neutrino oscillation in $\nu_e \to \nu_\mu$ channel what is puzzling.

The OPERA Collaboration observed a first candidate for $\nu_\tau$ appearance in the beam of mostly muon neutrinos produced at CERN and sent to Gran-Sasso [26]. This observation once confirmed with greater statistical significance would be a milestone in the neutrino oscillation physics.

## 2.4 Measurement of neutrino speed

Autumn of 2011 brought us the most unexpected result — the OPERA Collaboration performed a measurement of neutrino velocity accurately measuring the distance between production and detection points and synchronizing the clocks between CERN and Gran-Sasso with help of GPS to some nanoseconds (ns) level. As the result of analysis of 2009, 2010 and 2011 data neutrinos seem to arrive by about 60 ns earlier than expected for massless particle [27]. This result could be interpreted as a measurement of neutrino velocity which exceeds that of the light by about $2.5 \cdot 10^{-5} c$. At the moment of writing this short note more than one hundred of possible explanations or interpretations of this result and a lot of additional checks are already suggested by the community. Perhaps this result will not survive in the future. However it is already played an important role stimulating people to create new ideas, refresh the fundamentals of physics and re-asking the Nature again some «obvious» questions.

## 2.5 GSI anomaly

Another recent and puzzling result comes from GSI facility. It is not directly related to neutrino physics while some interpretations do make such a relation. GSI accelerator facility is used to study decays of highly ionized nuclei. Such ions can decay via weak interactions emitting neutrinos. GSI measures very accurately the life-time of some nuclei monitoring the trajectory of decaying nucleus including the trajectory of its dauther nucleus. The life-time distributions of Praseodymium

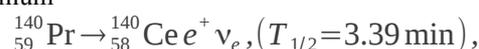
$${}^{140}_{59}\text{Pr} \to {}^{140}_{58}\text{Ce}\, e^+ \nu_e, (T_{1/2} = 3.39\,\text{min}),$$

and Promethium nuclei were studied:
$$^{142}_{61}\text{Pm} \rightarrow ^{142}_{60}\text{Nd}\, e^+ \nu_e, (T_{1/2}=40.5\,\text{sec}).$$

For both nuclei GSI observed an expected exponential distribution of the life-time with unexpected periodical modulation superimposed on top of it.

They fit the data by: $\frac{dN}{dT} = N(0) e^{-\lambda t} \lambda_{EC} (1 + a \cdot \cos(\omega t + \phi))$ and found a = 0.2 and ω=(7 sec)$^{-1}$ [28]. GSI suggested an interpretation of this puzzling result as a manifistation of emission of different massive neutrinos in the final state which according to their calculation should lead to quantum beats with period: $T = \frac{2M}{\Delta m^2} \approx 10\,\text{sec}$ if M = 140 $m_p$ and $\Delta m^2 = 10^{-4}\,\text{eV}^2$. This interpretation was questioned in the literature as it seem to violate causality or, perhaps saying it a bit more cautiously, this interpretation is in conflict with QM prescription that different final states should not interfere in the process amplitude. However, let us emphasize that a justification of this prescription was just a subject in the proposal of GSI studies.

Basically, they ask the following question. Consider an initial state $|i>$ which can end up in a final state $|f_k>$ with the corresponding amplitude $A(i \rightarrow f_k)$. What is the probability to observe any of the final states $|f_k>$ ? The QM prescription is well known: $|A(i \rightarrow f)|^2 = \sum_k |A(i \rightarrow f_k)|^2$. An alternative way to compute this probability would be the following formula $|A(i \rightarrow f)|^2 = \left|\sum_k A(i \rightarrow f_k)\right|^2$. Obviously, the last formula contains interference terms which are not present in the usual QM prescription. It is hard to find a really solid theoretical argument why QM prescription should be prefered therefore it sounds very reasonable to verify it experimentally. Thus whatever interpretation of the GSI anomaly would be accepted by the community in the future — it would be fair to say that GSI raised a really fundamental question.

## 3. Conclusions

26 years passed after the famous Pauli letter before the first anti-neutrino $(\bar{\nu}_e)$ was experimentally detected (1956). The second type of neutrino $(\nu_\mu)$ was discovered six years later (1962). Next 40 years were required to observe the third type of neutrino $(\nu_\tau)$ (2000). So, the first stage of neutrino study lasted for 70 years. Now the time is significantly compacted — new and very important results appear almost every year. This happens because of a large involvement of a wide community in neutrino physics. New and ambitious projects in neutrino physics are under active development. We bevelive that next years will bring us a lot of new results in neutrino physics.

## 3. Acknowledgements

It is my pleasure to thank the organizers of the Gamov School 2011 for inviting me to review neutrino physics at the School and for the warm atmosphere during the conference.